\documentclass[iop]{emulateapj}
\usepackage{natbib,graphicx,amsmath,xspace}

\newcommand{\mcg}{\mbox{MCG--5-23-16}\xspace}
\newcommand{\fluxu}{\mbox{ergs cm$^{-2}$ s$^{-1}$}\xspace}

\begin{document}

\title{Observations of \mcg with {\em Suzaku}, XMM-{\em Newton} and {\em NuSTAR}: Disk tomography and Compton hump reverberation}

\author{A. Zoghbi$^{1,2}$, E. M. Cackett$^3$, C. Reynolds$^{1,2}$, E. Kara$^4$ ,F. A. Harrison$^{5}$, A. C. Fabian$^{4}$, A. Lohfink$^1$, G. Matt$^{6}$, M. Balokovic$^5$, S. E. Boggs$^{7}$, F. E. Christensen$^{8}$, W. Craig$^{7}$, C. J. Hailey$^{9}$,  D. Stern$^{10}$, W. W. Zhang$^{11}$}
\affil{$^1$Department of Astronomy, University of Maryland, College Park, MD 20742-2421, USA}
\affil{$^2$Joint Space-Science Institute (JSI), College Park, MD 20742-2421, USA}
\affil{$^3$Department of Physics \& Astronomy, Wayne State University, 666 W. Hancock St, Detroit, MI 48201, USA}
\affil{$^4$Institute of Astronomy, Madingley Road, Cambridge CB3 0HA. UK}
\affil{$^5$Cahill Center for Astronomy \& Astrophysics, California Institute of Technology, Pasadena, CA 91125, USA}
\affil{$^6$Dipartimento di Matematica e Fisica, Universita degli Studi Roma Tre, via della Vasca Navale 84, I-00146 Roma, Italy}
\affil{$^7$Space Science Laboratory, University of California, Berkeley, California 94720, USA}
\affil{$^8$DTU Space. National Space Institute, Technical University of Denmark, Elektrovej 327, 2800 Lyngby, Denmark}
\affil{$^9$Columbia Astrophysics Laboratory, Columbia University, New York, New York 10027, USA}
\affil{$^{10}$Jet Propulsion Laboratory, California Institute of Technology, Pasadena, CA 91109, USA}
\affil{$^{11}$NASA Goddard Space Flight Center, Greenbelt, Maryland 20771, USA}

\email{azoghbi@astro.umd.edu}

\begin{abstract}
MCG--5-23-16 is one of the first AGN where relativistic reverberation in the iron K line originating in the vicinity of the supermassive black hole was found, based on a short XMM-{\em Newton} observation. In this work, we present the results from long X-ray observations using {\em Suzaku}, XMM-{\em Newton} and {\em NuSTAR} designed to map the emission region using X-ray reverberation. A relativistic iron line is detected in the lag spectra on three different time-scales, allowing the emission from different regions around the black hole to be separated. Using {\em NuSTAR} coverage of energies above 10~keV reveals a lag between these energies and the primary continuum, which is detected for the first time in an AGN. This lag is a result of the Compton reflection hump responding to changes in the primary source in a manner similar to the response of the relativistic iron K line.
\end{abstract}

\section{Introduction}
The primary X-ray continuum emission in AGN is due to Compton scattering of accretion disk photons in a hot corona \citep{1991ApJ...380L..51H}. The relatively cold and dense surrounding material intercepts and reflects some of this emission \citep{1991MNRAS.249..352G} producing a characteristic reflection spectrum \citep{1993MNRAS.261...74R,2013ApJ...768..146G}. Relativistic effects imprinted on the reflection spectrum provide a measure of the distance of the emitting region from the black hole allowing its spin to be directly measured \citep{1989MNRAS.238..729F,1991ApJ...376...90L,2006ApJ...652.1028B,2009ApJ...697..900M,2012ApJ...759L..15R}. The fast variability in the primary emission \citep[e.g.][]{2011MNRAS.413.2489V} is also mirrored by the reflecting medium (through a reverberation transfer function) providing direct measures of physical scales and geometries \citep{1992MNRAS.259..433M,1995MNRAS.272..585C,1999ApJ...514..164R,2013MNRAS.430..247W}.

After a hint of detection in Ark 564 \citep{2007MNRAS.382..985M}, reverberation lags were significantly observed in several Seyfert galaxies, where the soft (0.5-1 keV) band, composed mostly of reflected iron L emission and other species, is observed to lag the harder (1-4 keV) band, which is dominated by the primary emission \citep{2009Natur.459..540F,2010MNRAS.401.2419Z,2011MNRAS.418.2642Z,2011MNRAS.416L..94E,2011MNRAS.417L..98D,2013MNRAS.429.2917F,
2013MNRAS.430.1408K,2013ApJ...764L...9C}. The time delay and the time-scale on which it is observed both appear to scale with black hole mass \citep{2013MNRAS.431.2441D}. 

A major step forward was the discovery of iron K reverberation in NGC 4151 \citep{2012MNRAS.422..129Z}. Whereas the earlier lags were between the primary continuum and a blend of relativistically-broadened emission lines (comprising the soft excess), the lags in NGC 4151 were between the same primary continuum and the iron K line at $\sim 6.4$ keV. The significance of the iron K reverberation is that it is a single line in a clean part of the spectrum, allowing direct spectro-timing modeling that directly probes the emitting region \citep{2014MNRAS.438.2980C,2014MNRAS.tmp..449E}. Iron K line reverberation has now been observed in several other objects \citep{2013ApJ...767..121Z,2013MNRAS.434.1129K,2013MNRAS.tmpL.209K,2014arXiv1402.7245M}, and again the delay and the time-scale on which it is observed, both appear to scale with mass \citep{2013MNRAS.434.1129K}.

\begin{figure*}
\centering
 \includegraphics[width=410pt,clip ]{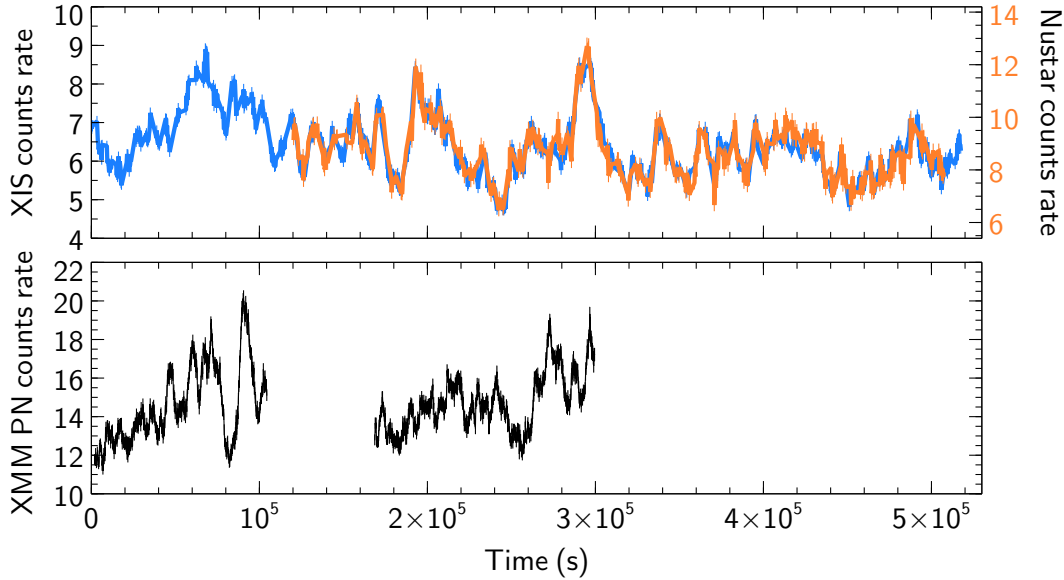}
\caption{2--10 keV light curves from all the new observations of \mcg. \emph{Top:} {\em Suzaku} XIS (combined front and back-illuminated) from two observations (obs ids: 708021010 and 708021020), shown in blue (left axis). The panel also show in orange (right axis) the simultaneous {\em NuSTAR} light curve (obs id: 60001046002), where data from FPMA and FPMB have been combined. \emph{Bottom:} XMM-{\em Newton} EPIC PN light curves from the full two orbits (obs ids: 0727960101 and 0727960201, not simultaneous with top panel) taken three weeks later.}
\label{fig:lc}
\end{figure*}

In this work, we study the reverberation in \mcg in detail and present the first measurement of a lag in the Compton hump in an AGN. A new observation of the source with the Nuclear Spectroscopic Telescope Array \citep[{\em NuSTAR};][]{2013ApJ...770..103H} revealed a time delay above 10 keV that is consistent with reverberation in the innermost regions of the accretion disk.

\mcg is a Seyfert 1.9 Galaxy ($z=0.0085$) with a typical 2--10 keV flux of $\sim8\times10^{-11}$ \fluxu and a mass of $1-5\times10^7$ M$_{\odot}$ (\citealt{1986ApJ...306L..61W, 2012A&A...542A..83P}). Its spectrum resembles a classical Compton-thin Seyfert 2 galaxy with a column that does not affect the spectrum above 3 keV significantly. The spectrum below 1 keV contains a combination of emission from scattered continuum photons and distant photoionized gas. Above 2 keV, the spectrum shows both narrow (EW$\sim$60 eV) and broad (EW$\sim50-200$ eV) iron K$\alpha$ lines along with a strong Compton hump above 10~keV (\citealt{1997ApJ...474..675W,2004A&A...415..437B, 2004ApJ...601..771M, 2006AN....327.1067B, 2007PASJ...59S.301R,2013ApJ...767..121Z}).

This work uses new observations of \mcg with {\em Suzaku}, XMM-{\em Newton} and {\em NuSTAR} with three specific aims:
\begin{itemize}
\item To confirm the iron K reverberation lags reported in \cite{2013ApJ...767..121Z} using a single XMM-{\em Newton} observation.
\item To use the combined datasets to measure lag-energy spectra on three time-scales, localizing the emission to three emission regions
\item To observe time lags due to reverberation in the Compton reflection hump using {\em NuSTAR}.
\end{itemize}

\section{Data Reduction \& Analysis}
In order to study the detailed variability of \mcg, it was targeted with a long {\em Suzaku} observation starting on June, 01 2013 for a net exposure of 300 ks (obs ids: 708021010 and 708021020). {\em NuSTAR} observed it simultaneously starting on June 03, 2013 for an exposure of 150 ks (obs id: 60001046002). XMM-{\em Newton} \citep{2001A&A...365L...1J} also observed \mcg for two full orbits three weeks later starting on June 24, 2013 (obs ids: 0727960101 and 0727960201). We concentrate in this work on the timing results. Spectral modeling will be published separately.

{\em Suzaku} and {\em NuSTAR} data were reduced using \textsc{heasoft v6.15} with the latest calibration files (CALDB files released December 23, 2013 and February 3, 2014 for {\em NuSTAR} and {\em Suzaku} respectively). XMM-{\em Newton} data were reduced using \textsc{sas} (\emph{xmmsas\_20131209\_1901-13.5.0}) and also using the latest \textsc{ccf} calibration files including the EPIC-pn \citep{2001A&A...365L..18S} Long-Term charge transfer inefficiency (CTI) corrections. {\em Suzaku} light curves were extracted from different energies (section \ref{results}) using \textsc{xselect} to filter events from a source and background regions. Background counts scaled to the source region size were then subtracted from the source counts. {\em NuSTAR} light curves for source and background regions were corrected for livetime, psf, exposure and vignetting using the task \textsc{nulccorr}. Absolute and relative corrections for the EPIC PN and MOS light curves were applied using \textsc{epiclccorr}. For the purpose of the analysis presented here, all light curves have time bins of 512 seconds. Other time bins do not change the results.

Time lags are calculated using the method detailed in \cite{2013ApJ...777...24Z}. The method uses maximum likelihood to obtain the frequency-resolved time lags by directly fitting the light curves \citep{2010MNRAS.403..196M}. The uncertainties in the lag measurements represent the 68\% confidence intervals calculated as the values that change $-2log(\mathcal{L}/\mathcal{L}_{max})$ by 1 where $\mathcal{L}$ is the likelihood value and $\mathcal{L}_{max}$ is the optimum likelihood value \citep[see details in ][]{2013ApJ...777...24Z}. This means the lag errors in the plots represent 1$\sigma$ uncertainties. The errors quoted in the text when doing model fitting represent the $90\%$ confidence limits for one interesting parameter (i.e. $\Delta\chi^2=2.71$).

\section{Results}\label{results}
\subsection{Iron K reverberation}\label{sec:ironK_rev}
In this section we concentrate on the $2-10$ keV energy band. Here we use the combined XMM-{\em Newton}, {\em Suzaku} and {\em NuSTAR} ($<10$ keV) datasets to study time lags in the iron K line. We first present a simple method for directly observing lags in the light curve. We then present a more general analysis using frequency-resolved time lags. 

\subsubsection{Simple considerations}\label{sec:simple_kline}
For a start, we consider a simple procedure to study the lag as a function of energy. The light curves from the new observations are plotted in Fig. \ref{fig:lc}. The top panel shows the simultaneous {\em Suzaku} XIS (longest blue) and {\em NuSTAR} light curves (orange), while the XMM-{\em Newton} EPIC-PN light curve is plotted in the lower panel. The light curves are characterized by high variability and strong flares.

\begin{figure}
\centering
\begin{tabular}{cc}
 \includegraphics[width=95pt,clip ]{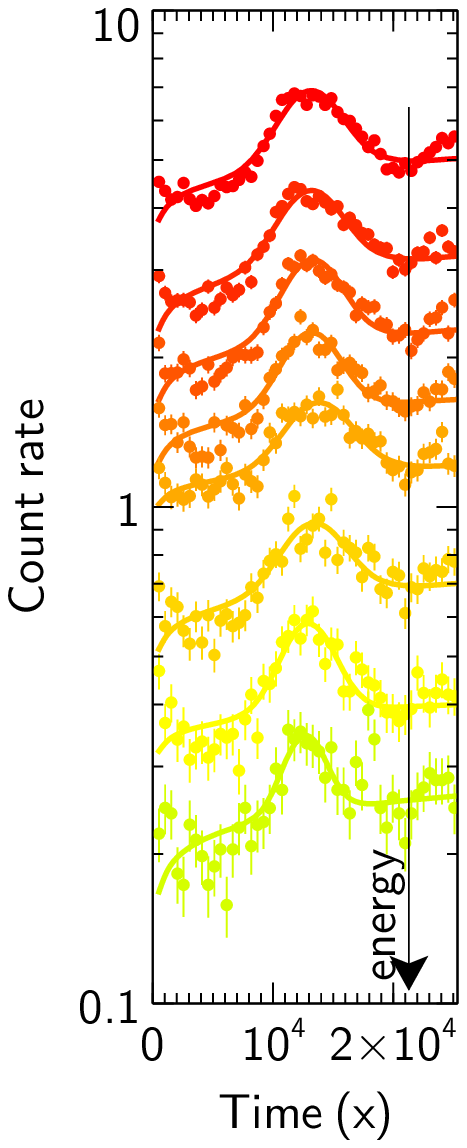} &
 \includegraphics[width=133pt,clip ]{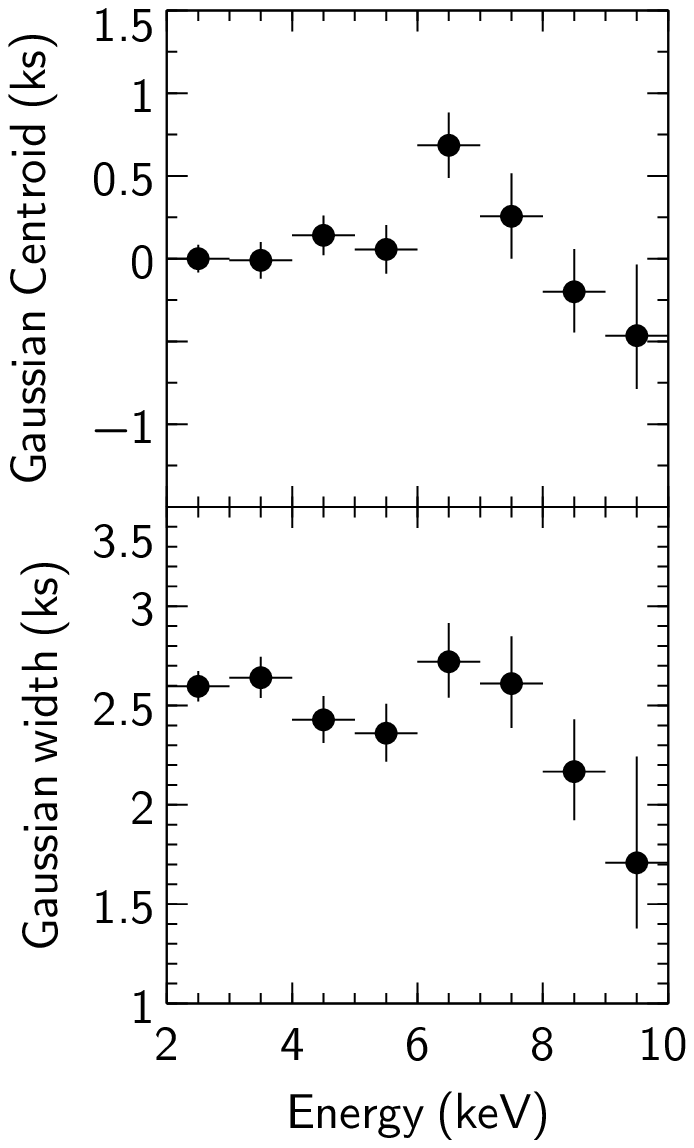}
\end{tabular}
\caption{\emph{Left:} A plot of the light curve section showing the strong flare at the end of the first XMM-{\em Newton} light curve plotted at different energies (energy increases down, and the x-axis has been shifted to zero). The lines on top of each flare represent the best fitted Gaussian model (see section \ref{sec:simple_kline}). \emph{Right-top:} The best fit values for the peak of each Gaussian model from the left panel (i.e. the lag of each flare compared to the 2--3 keV flare) plotted as a function of energy. There is a clear feature at the 6--7 keV band. \emph{Right-bottom:} The best fit values for the width of each Gaussian model from the left panel. The width of the flares also changes with energy, and tracks the shape of an iron line similar to the flare peak.}
\label{fig:2480_fit}
\end{figure}

It is known from a previous XMM-{\em Newton} observation \citep{2013ApJ...767..121Z} that the peak of the iron line ($\sim 6.4$~keV) is delayed with respect to lower and higher energies on time scales longer than $\sim 10 ks$. The simplest test one can do is look for lags within flares of a particular time scale at different energies. Although most flares contain multiple time scales (i.e. small scale flares on top of a longer time scale trend), the single flare at the end of the first XMM-{\em Newton} orbit (just below $10^5$ s in Fig. \ref{fig:lc}-bottom) is particularly interesting. It appears mainly as a $\sim 10$ ks flare. 

Fig. \ref{fig:2480_fit}-left shows a zoom in of this flare from eight energy bins between 2 and 10 keV in steps of 1 keV. The flare in each energy band was fitted with a simple Gaussian model to obtain a measure of the peak of the flare. The best fitted Gaussian centroids and their widths  are plotted in Fig. \ref{fig:2480_fit}-right. The best-fit Gaussian peaks were shifted vertically with the same amount so that the first point has zero lag. It is clear from the change of the peak of the flare that the iron line lags the bands above and below it. The shape of the `broad iron line feature' is strikingly similar to that obtained from full Fourier analysis of the previous observation \citep{2013ApJ...767..121Z} despite the fact that it is extracted only from a single flare. Fig. \ref{fig:2480_fit}-right also shows the width of the flare as a function of energy. Interestingly, this too has a broad-line-like feature peaking at 6--7 keV.

\begin{figure}
\centering
 \includegraphics[width=250pt,clip ]{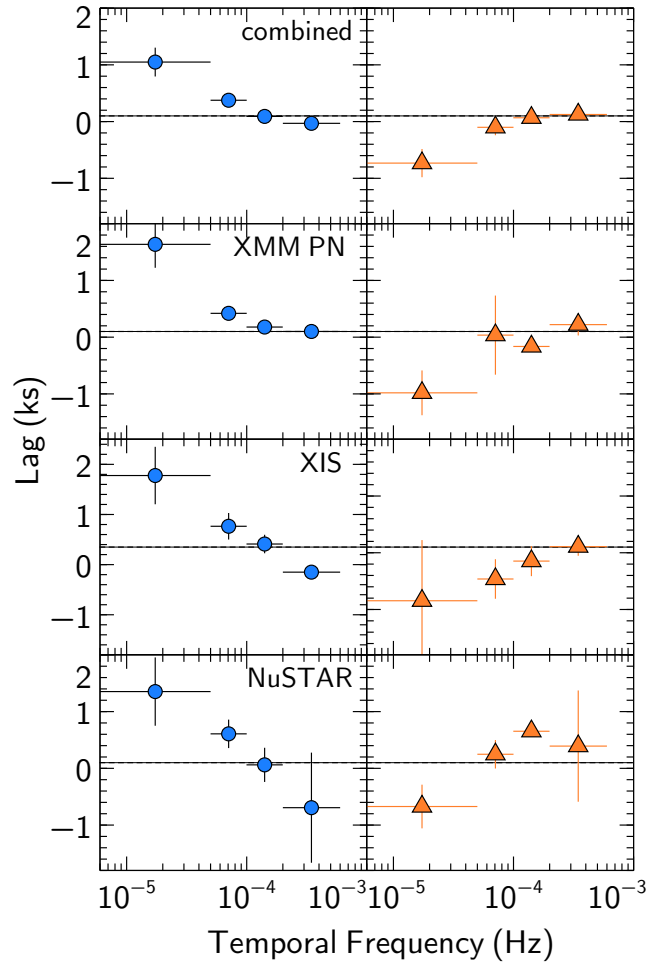}
\caption{Frequency-resolved time lags for \mcg comparing the peak of the iron K line to its wings from combined analysis (first row) and from individual instruments (rows 2--4, as labeled). The left column compares the 6--7 keV band to the 2--3 keV band (blue circles). The right column (orange triangles) compares the 6--7 keV band to the 9--10 keV band (for XIS, we use 7--8 keV instead of 9--10 keV to obtain a better signal). We follow the usual convention, where positive means hard lags, i.e. hard band lagging soft band. The left column shows that the 6--7 keV band \emph{lag} bands \emph{below} it (positive lag). The right column shows that the 6--7 keV lags bands \emph{above} it (negative lags). The different rows show that this conclusion is consistent between the different instruments used.}
\label{fig:lag_fq}
\end{figure}

\begin{figure*}
\centering
 \includegraphics[height=200pt,clip ]{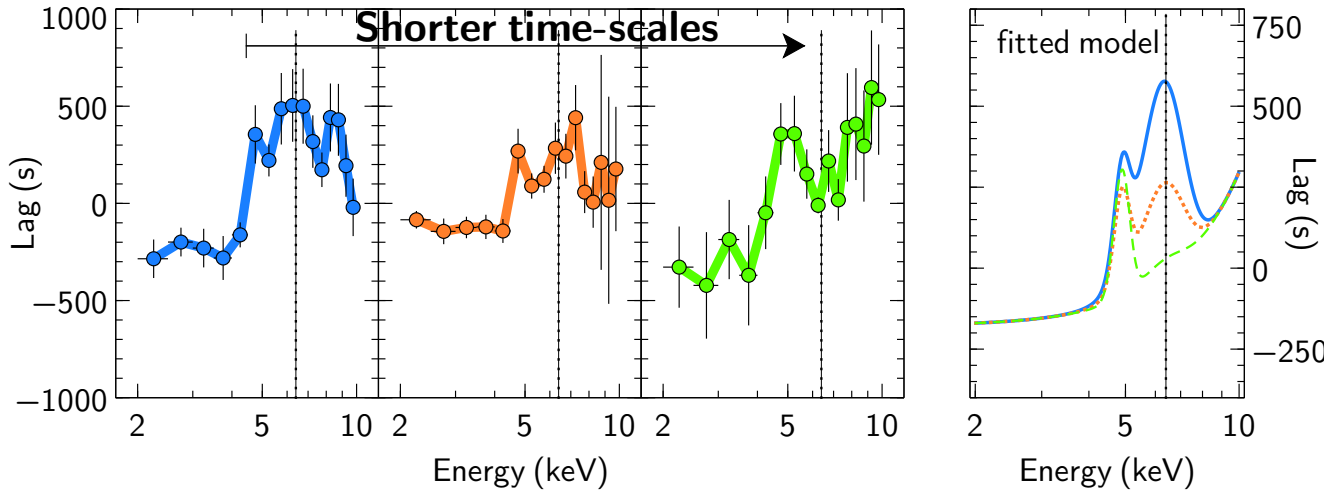}
\caption{\emph{Left:} Lag -energy spectra for \mcg at three frequency bins from a joint fitting of the XMM-{\em Newton}, {\em Suzaku} and {\em NuSTAR} light curves. From left to right: $2\times10^{-5}-10^{-4}$ Hz, $10^{-4}-2.5\times10^{-4}$ Hz and $2.5\times10^{-4}-5\times10^{-4}$ Hz. \emph{Right: } The best-fit model for the three frequencies consisting of two Gaussian lines. Solid (blue), dotted (orange) and dashed (green) lines correspond to the three frequency bins in Fig. \ref{fig:lag_en} in increasing frequency respectively. The normalization of the first Gaussian peaking at $\sim 5$ keV remain unchanged within statistical uncertainties, while the second Gaussian line peaking at $\sim 6.5$ keV decreased in strength as we move to higher frequencies.}
\label{fig:lag_en}
\end{figure*}

\subsubsection{Frequency-resolved time lags}
The previous method does not take into account the stochastic nature of the light curve. Therefore, in order to study the lags in a systematic manner, using all the available data, we explore the frequency-resolved time lags. Although the XMM-{\em Newton} PN light curves are evenly sampled, those from {\em Suzaku} and {\em NuSTAR} are not due to Earth occultations. We use a maximum likelihood method to take this into account by directly fitting for the time lags at different temporal frequencies. The method divides the observed frequency range into a number of bins taking the lag value at each bin as an unknown parameter. A covariance matrix is constructed by an inverse Fourier transform and is used to calculate a likelihood function. Assuming uniform priors on the parameters and using a Bayesian formalism, lags at different frequencies and their probability distribution are obtained \citep[see][]{2013ApJ...767..121Z}. The end result when comparing two light curves is a lag-frequency plot. If more than one light curve is available (e.g. from different energies), each light curve is compared to a reference light curve, and the lag-energy plot can be constructed for the temporal frequencies of interest.

Fig. \ref{fig:lag_fq} shows the lag-frequency plot for the peak of the iron line (6--7 keV band) compared to both lower (2--3 keV; left column) and higher energies (right column). Light curves from the different detectors were fitted together by constructing a joint likelihood as well as separately. Positive and negative values indicated hard (i.e. hard bands lag soft bands) and soft lags respectively. Two clear results can be seen in Fig. \ref{fig:lag_fq}:
\begin{itemize}
\item The 6--7 keV band lags both lower energies (hard lags in the left panel) and higher energies (soft lags in the right panel).
\item Different detectors show the same trends where the 6--7 keV band lags both lower and higher energy bands.
\end{itemize}
This result is a generalization of the phenomenon shown in Fig. \ref{fig:2480_fit} shows for a single flare in the XMM-{\em Newton} light curve. It further confirms what has been seen in previously in shorter observations, where the the iron line peaking at 6--7 keV lags the primary continuum that dominates energies at either side of the line.

To further study these lags, we present the energy-dependent lags. The previous study of \mcg showed that the lag-energy profile averaged over a relatively broad frequency band resembles a broad iron line. Our aim here is try to probe this profile for several frequency bands (i.e. different time scales).

\subsubsection{Detailed energy-dependence of the lag}
Light curves in sixteen energy bins between 2 and 10 keV in steps of 0.5 keV were extracted. This was found to be an optimum trade-off between energy resolution and having reasonable uncertainties in the calculated lags. The lag of the light curves in each energy band was calculated taking the whole 2--10 keV band as a reference (excluding the band of interest each time so the noise remains uncorrelated). The frequency band where there is high signal was divided into three bins. The resulting lag-energy profiles are shown in Fig. \ref{fig:lag_en}-left.

The lag-energy profiles are plotted for three frequency bands: $2\times10^{-5}-10^{-4}$, $10^{-4}-2.5\times10^{-4}$ and
$2.5\times10^{-4}-5\times10^{-4}$ Hz. For reference, the first XMM-{\em Newton} results in \cite{2013ApJ...767..121Z}
were for frequencies $<10^{-4}$ Hz, which corresponds to the first frequency bin. The results presented here and those
in \cite{2013ApJ...767..121Z} are remarkably similar despite the different observations, confirming that the observed
delays are not transients, but rather a generic property of the object. The lowest frequency in the plots is
$2\times10^{-5}$ to a direct comparison with the \citealt{2013ApJ...767..121Z} can be made. Including the lowest
available frequencies makes the lags slightly larger (it can be seen in Fig. \ref{fig:lag_fq}) but the shape remain the same.

The plots at the three frequency bins show clear line profiles. The shape of line appears to change with temporal frequency, peaking towards lower energies for the highest frequency bin (shortest time scales). The data above $\sim 5\times10^{-4}$ Hz are dominated by noise, and the lag-energy spectrum (not shown) is consistent with constant zero lag. The lag plot for all frequencies is the average of the three profiles, and it is dominated by the lowest frequency band. The energy profile in this case looks very similar to the lowest frequency profile.

In all three spectra shown in Fig. \ref{fig:lag_en}-left, a constant or a smoothly increasing lag is ruled out with very high significance. Fitting these profiles with a simple model of a power-law and a Gaussian to characterize the existence and the general shape of the lag, we obtain the parameters shown in Table \ref{tab:fit} ($\chi^2=49$ for 33 degrees of freedom). Although the Gaussian energies in the lowest two frequency bins differ only at the $\sim 80\%$ confidence level, the highest frequency bin is significantly ($>99.999\%$ confidence) shifted to lower energies.

If instead we fit the spectra with a model consisting of two Gaussian lines, we find the best fitting model shown in Fig. \ref{fig:lag_en}-right. This model is motivated by the appearance of the lag profiles. In particular, the rise at $\sim 4$~keV appears to be persistent at all frequencies, while the lag value at $6-7$~keV changes significantly. This model describes the data well with a $\chi^2=39$ for 35 degrees of freedom. The best fit values for the Gaussian lines are all consistent within a 1$\sigma$ uncertainty, except for the high energy Gaussian at the highest frequencies where it is not required. Therefore, we assumed the Gaussian lines to have the same energies and widths in all three frequency bins. This assumption allows the three bands to be directly compared.

\begin{table}[h]
\centering
\caption{One-Gaussian model fit parameters.}
\begin{tabular}{llll}
\hline\hline
Parameter&Bin 1 & Bin 2 & Bin 3 \\ \hline
$E$ (keV) & $7.06\pm 0.42$ & $6.62\pm0.45$ & $5.03\pm0.24$ \\
$\sigma$ (keV) & $2.13\pm 0.42$ & $1.29\pm0.42$ & $0.49\pm0.2$ \\
\hline
\end{tabular}
\label{tab:fit}
\end{table}

Although the model in Fig. \ref{fig:lag_en}-right is simply a descriptive and not physical model, it provides valuable
insights in understanding these lag-energy spectra. The low energy Gaussian line peaks at $4.9\pm0.5$ keV ($90\%$
confidence), and does not appear to change with temporal frequency (all normalization values are within one-sigma
uncertainty). The high energy Gaussian on the other hand, peaks at $6.28\pm0.14$ keV, and its strength decreases
gradually towards higher temporal frequency (normalization: $1234\pm280$, $589\pm223$ and $106\pm120$ s keV for the three frequency bins in increasing frequency respectively, see Table \ref{tab:fit2}).

\begin{table}[h]
%\centering
\begin{center}
\caption{Two-Gaussian model fit parameters.}
\begin{tabular}{llll}
\hline\hline
Parameter&Bin 1 & Bin 2 & Bin 3 \\ \hline
$E_1^{\ast}$ (keV) & $4.9\pm 0.5$ & -- & -- \\
$norm_1$ (s keV) & $174\pm78$ & $160\pm59$ & $225\pm80$ \\
$E_2^{\ast}$ (keV) & $6.28\pm 0.14$ & --& -- \\
$norm_2$ (s keV) & $1234\pm 280$ & $589\pm223$ & $106\pm120$ \\
\hline
\end{tabular}
\label{tab:fit2}
\end{center}
$^{\ast}$ The energy was fixed between the three bins.
\end{table}

If we further assume that the line profiles are due to a relativistic iron line, we can replace the two-Gaussian model with a relativistic line model. This does not take into account the full relativistic transfer function, but provides a tool towards understanding these lag-energy profiles. We use the \textsc{laor} \citep{1991ApJ...376...90L} model, which is a delta function convoluted with a kernel of the Kerr metric. Its parameters in addition to the line energy and normalization are the inner and outer radii of emission ($r_{in},r_{out}$), emissivity index $q$ and inclination. The model also fits the data reasonably well ($\chi^2/$d.o.f$=61/40$). Individual spectra could not constrain all the model parameters separately, so we made the assumption that most parameters are the same for all frequency bins. $r_{out}$ was left as a free parameter and it is used as a proxy for the extent of the emission region. This will be the case when, to first order, the variability time scale (i.e. temporal frequency) is directly associated with the emission region size. The inclination of $38$ degrees and emissivity index of 2.2 from the spectral modeling in \cite{2013ApJ...767..121Z} were used.

\begin{figure}
\centering
 \includegraphics[height=130pt,clip ]{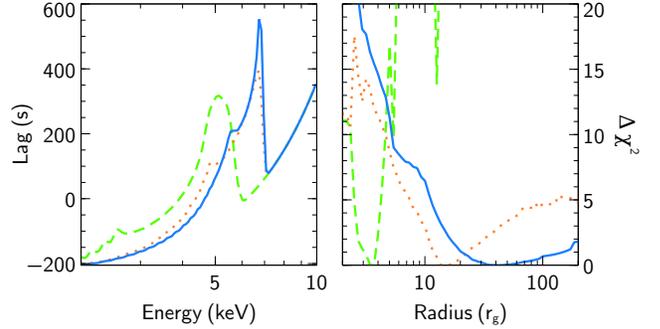}
\caption{\emph{Left: } The best-fit model to the lag-energy spectra. The model consists of a power-law and a relativistic \textsc{laor} line. Solid (blue), dotted (orange) and dashed (green) lines correspond to the three frequency bins in Fig. \ref{fig:lag_en} in increasing frequency respectively. \emph{Right: } Confidence levels for the outer radius of emission plotted as $\Delta\chi^2$ difference from the best fit $\chi^2$ value.}
\label{fig:laor}
\end{figure}

The results of the fit are shown in Fig. \ref{fig:laor}. The left panel shows the model for the three frequencies, while the right panel shows the confidence ranges for the outer radius parameter plotted as the $\Delta\chi^2$ difference from the minimum $\chi^2$ value. The figure clearly shows how the extent of the emission region inferred from the lag-energy profiles changes with temporal frequency, or equivalently with variability time scale. At the longest time scales (solid blue line in Fig. \ref{fig:laor}), the emission region extends over a larger distance ($>50 r_g$), and it becomes smaller when shorter time scales are probed ($<5 r_g$ for the shortest time scale).

\subsection{Compton Hump reverberation}\label{sec:hump_lags}
After discussing the K-band reverberation ($2-10$ keV band), we present in this subsection the result of analyzing the {\em NuSTAR} light curves above 10 keV. Following similar steps to those used when discussing the iron K band, we start with the simple procedure then we consider a full frequency-resolved case.

\subsubsection{Simple consideration}\label{sec:hump_simple}
\begin{figure}
\centering
 \includegraphics[width=240pt,clip ]{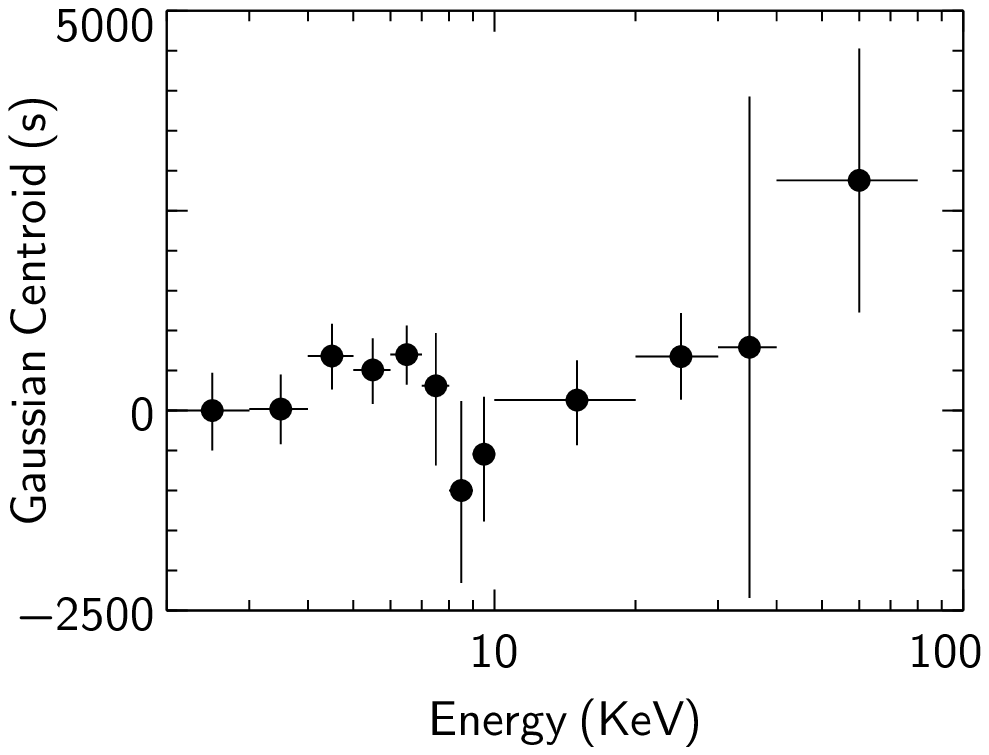}
\caption{The peak of the light curve flare at different energies from the {\em NuSTAR} light curve. All the times are calculated relative to the first point. This is similar to the XMM-{\em Newton} example shown in Fig. \ref{fig:2480_fit}-right. See section \ref{sec:hump_simple}.}
\label{fig:nu_fit}
\end{figure}

The method in this section is similar to that in \ref{sec:simple_kline}, where we fit a simple Gaussian line to an isolated flare in the light curve. We select the strongest flare in the {\em NuSTAR} light curve ($\sim2.9\times10^{5}$ seconds in Fig. \ref{fig:lc}) in twelve energy bands between 2 and 80 keV. Light curves from the two modules FPMA and FPMB were combined. A plot of the flare peak time as a function of energy is plotted in Fig. \ref{fig:nu_fit}. All the points were shifted vertically with the same amount so that the first point is zero.

Although this procedure is again simplistic and does not take into account the stochastic nature of the light curves, it provides some useful insights. The peak of the flare seems to change with energy. Below 10 keV, the flare shifts are consistent with those in the XMM-{\em Newton} data and with the full frequency-resolved analysis of section \ref{sec:ironK_rev}. The data above 10 keV show an increase with energy. Only one flare is used to produce Fig. \ref{fig:nu_fit}, and the features are statistically not very significant (a constant null hypothesis is not ruled out.). The following section studies the lag-energy profile systematically using full Fourier-resolved time lags.

\subsubsection{Fourier-resolved time lags}
\begin{figure*}
\centering
 \includegraphics[width=420pt,clip ]{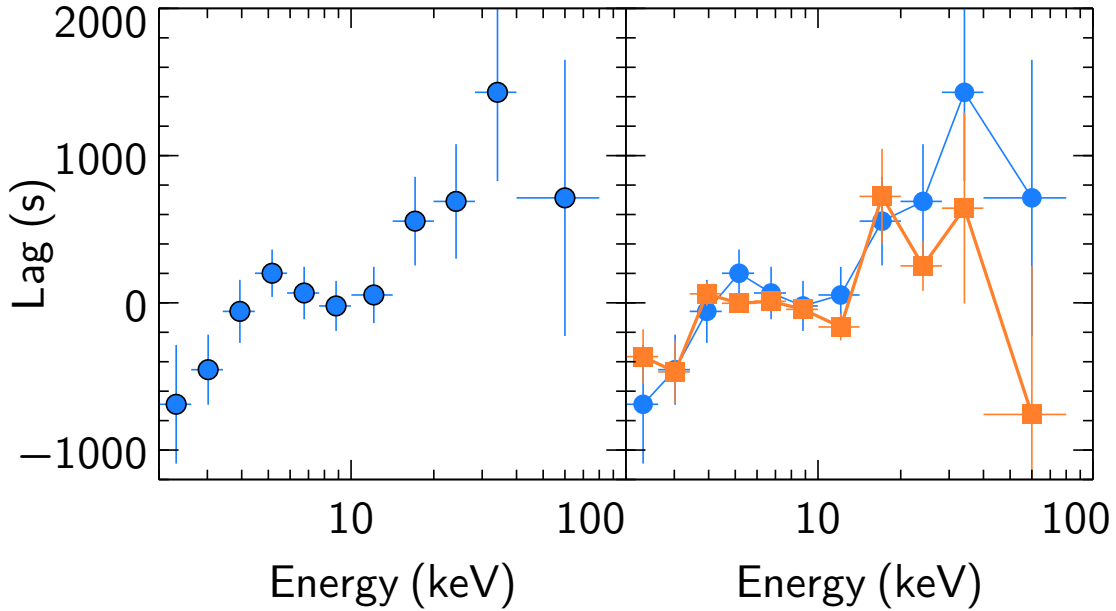}
\caption{Lag-energy spectra for \mcg using {\em NuSTAR} light curves. \emph{Left: } Lag-energy spectra for lags averaged over a wide frequency band covering $6\times10^{-6}-6\times10^{-4}$ Hz. \emph{Right: } Lag-energy spectra at two frequency bands. Blue circles are for the same frequency band as the left panel. The orange squares are for frequencies $3\times10^{-5}-6\times10^{-4}$ Hz. The central frequencies for the two bins are: $6\times10^{-5}$ (blue circles) and $10^{-4}$ Hz (orange squares) respectively.}
\label{fig:nu_lag_en}
\end{figure*}

The left panel of Fig. \ref{fig:nu_lag_en} shows the lag-energy spectrum resulting from Fourier-resolved time lag analysis using the {\em NuSTAR} data. The light curves from the two modules FPMA and FPMB are fitted simultaneously. The left plot is for lags averaged over the whole frequency range covering $6\times10^{-6}-6\times10^{-4}$ Hz. The light curves were divided into segments of length $\sim30-40$ ks with a 512 seconds time sampling. Eleven energy bins with logarithmic binning were chosen, 6 below 10 keV and 5 above it.

Fig. \ref{fig:nu_lag_en}-left shows a general increase in the lag with energy with the $\sim 6$ keV feature that is consistent with what is seen in the combined data analysis of section \ref{sec:ironK_rev}. There is also an increase above 10 keV with a possible peak at $\sim 30$ keV. The increase above 10 keV is very significant (e.g. against a constant null hypothesis). The feature at $\sim 6$ keV is significant at $>97\%$ confidence when fitted with a Gaussian against a null hypothesis of a single power-law for the whole spectrum (this increases to more than $99.9\%$ if the energy of the line is assumed known from photon spectral modeling for example). The turnover at 40 keV is significant at the $\sim 90\%$ confidence when comparing a power-law fit with and without a cutoff energy.

If our interpretation of the feature at 6 keV as due to relativistic reverberation in the iron K line (section \ref{sec:ironK_rev}) is correct, then a reverberation in the Compton hump is expected above 10 keV. If the iron line is responding to the primary continuum variability, then so does the Compton hump produced as part of the reflection spectrum. The effect of a possible intrinsic continuum lag (i.e. lags in the primary X-ray source that has nothing to do with reflection) are discussed in section \ref{sec:discussion}.

To investigate the frequency-dependence of the lag-energy spectrum in the {\em NuSTAR} data, Fig. \ref{fig:nu_lag_en}-right shows the same plot as in the left panel, and also a lag-energy spectrum at the highest frequencies ($>3\times10^{-5}$ Hz). This is slightly different from the plots in Fig. \ref{fig:lag_en}. There, the signal was high enough to extract lag-energy spectra in three adjacent and independent frequency bins. Here, the two frequency bins overlap above $3\times10^{-5}$ Hz, but they probe different ranges. A combination of large scatter and large uncertainties dominate the lag-energy spectra when independent frequency bins are considered. The central frequencies for the two bins considered in Fig. \ref{fig:nu_lag_en} are: $6\times10^{-5}$ (blue circles) and $10^{-4}$ Hz (orange squares) respectively.

Fig. \ref{fig:nu_lag_en}-right shows two key features. First, the lag-energy plot appears to be generally steeper at lower frequencies, due possibly to the contribution from intrinsic lags (see discussion in section \ref{sec:discussion}). Second, the two features at 6 and 40 keV appear to shift to lower energies at higher frequencies. The first statement is significant at more than the $4\sigma$ level (for example we fit a linear model to the log of the energy we find slopes of $408\pm118$ and $75\pm64$ for the low and high frequencies respectively), while the second statement is significant at $\sim 2\sigma$ level (when fitted with two Gaussians, one for each peak, the energies shift from $5.2\pm0.5$ and $42\pm6$ keV at low frequencies to $4.3\pm0.7$ and $32\pm4$ keV at high frequencies for the low and high energy Gaussians respectively).

\section{Discussion}\label{sec:discussion}
This work presents an analysis that confirms the existence of reverberation lags in \mcg. The lags are detected in new observations with XMM-{\em Newton}, {\em Suzaku} and {\em NuSTAR}. Two further main results are presented: 1- The detection of a time-scale dependent change in the shape of the broad iron K line seen in the lag-energy spectra. 2- The detection of a time delay in the {\em NuSTAR} light curves that is consistent with reverberation in the Compton reflection hump.

\subsection{Reverberation below 10 keV: Disk tomography}

The basic conclusion from the first result is that time lags seen in the short XMM-{\em Newton} data in \cite{2013ApJ...767..121Z} are not only confirmed, but the new observations allow the emission from different regions to be separated. The results in the lag-energy spectra are \emph{model-independent} as they are merely the outcome of comparing time shifts and delays between energy bands. The delays themselves have to be associated with a reflection process because there is a clear signature at the iron K energy at $\sim 6.4$ keV.

The primary continuum as well as the reflection both contribute at all energies between 2 and 10 keV, but with different fractions at different energies depending on the shape of the reflection spectrum \citep{1991MNRAS.249..352G,1993MNRAS.261...74R}. As discussed in \cite{2001AdSpR..28..267P} \citep[see also][]{2013ApJ...767..121Z}, the lag-energy spectra measure the relative contribution of the reflection compared to the primary (power-law) component. In other words, the lag-energy profile measures the reflection fraction as a function of energy. For a simple spectral model consisting of a single power-law for the primary continuum and a reflected iron K line, as is the case here, the lag-energy spectrum roughly traces the shape of the iron K line itself. The measured shape for \mcg is clearly not a narrow line, indicating that relativistic effects are important.

The analysis in section \ref{sec:simple_kline} shows that by simply comparing flares at different energies, we find that the peak of a broad line lags energies below and above it. The widths of the flare changes accordingly. In a reverberation picture, the widths should be narrowest for the energies where the primary continuum dominates, and increase with increasing refection fraction.  This is a simple expectation of a reverberating signal, where the reflected signal is a lagged, \emph{smeared}, version of the irradiating signal, with the amount of smearing increasing with the size of the reverberating region.

There is always a lag dilution factor caused by both primary and reflected components contributing to the energy bands of interest. The amount of dilution depends on the shape of the photon spectrum \citep[e.g.][]{2013MNRAS.430..247W}. Although the spectrum of the source in the observations used here will be discussed in detail in a separate work, the general shape is not very different from that reported in \cite{2013ApJ...767..121Z} for the first XMM-{\em Newton} observation. The lags between bands measured here are of order hundreds to thousands seconds. The reflection fraction changes from $\sim30\%$ at the peak of the relativistic line to $\sim10-20\%$ at the wings. This gives reflection-to-primary lags of order a few  to several kilo-seconds. The light crossing time in seconds at a distance $r$ (in units of $r_g=GM/c^2$) from a black hole is $GMr/c^3 = 50M_7r$ where $M_7$ is the black hole mass in units of $10^7M_{\odot}$. The mass of \mcg is not known accurately, but it is of the order $M_7\sim1-5$ \citep[e.g.][]{2012A&A...542A..83P}. Therefore, a delay of a few kilo-seconds between the primary and the reflection, to within a factor accounting for the geometry, corresponds to a few to a few tens of gravitational radii.

Interpreting the lag-energy spectra at different frequencies starts with the assumption that there is a correspondence between time-scale and emission region size. Under this picture, different time-scales (i.e. temporal frequencies) probe different emission region sizes. Combining this with the fact that lag-energy spectra measure the shape of the reflection component relative to the primary continuum that has a power-law shape, one can hope to measure the shape of the relativistic iron line emitted in different regions. The plot in Fig. \ref{fig:lag_en} shows this principle. The rest energy of the iron K line is  6.4 keV. The lag-energy spectrum at the lowest frequencies is broad and peaks at this energy, while the spectrum at higher frequencies peaks at lower energies. The line profile of a relativistic line from an accretion disk is well understood.
Photons in the red wing of the line are emitted deep in the black hole potential potential, and they are emitted closer to the black hole.Photons emitted further from the black hole on the other hand, have a shallower black hole potential to escape and they are observed closer to the rest energy of the line.

An important point to note from the energy and frequency lag dependence is the fact that the red wing of the lag profile is the same at all time scales, while the core is more apparent at the longest time scales only, as expected from Fe K reverberation.  As described in \citealt{2014MNRAS.438.2980C}, the higher frequencies select against the longest lags producing a cut-out region in the core of the line that increases from low to high frequencies.

The lag magnitude at the rest energy of the line changes from $\sim700$ seconds at long time scales to $0-200$ seconds at the highest frequencies (Fig. \ref{fig:lag_en}). This factor of $\sim4$ change in the lag corresponds to a factor of $\sim7$ change in time-scale (the difference in frequency between the first and last bin in Fig. \ref{fig:lag_en}), which is consistent knowing that the light crossing time (which scales with the lag) is proportional to the radius $r$ while the time-scale (most likely thermal in this case) scales with $r^{3/2}$ \citep{2002apa..book.....F}. We attempted to do a full modeling of the lag-energy spectra at the three frequencies using the simple lamp-post model in \cite{2014MNRAS.438.2980C}. We found an excellent fit to the lowest two frequencies with a source height of $h\sim10 r_g$. The model was unable to produce the highest frequency shape indicating possibly that the assumed geometry is not correct, and the corona is likely extended vertically. The observation seems to indicate that the shortest time-scale response has to be emitted only at small radii illuminated by possibly a different source than the longest time-scales. We defer exploring models with different illuminating geometries to a future work.

\subsection{Reverberation in the Compton hump}
The analysis in section \ref{sec:hump_lags} presents the first measurement of a time delay above 10 keV in an AGN. This is only possible thanks to the unprecedented hard X-rays sensitivity provided by {\em NuSTAR}. The existence of the lags in the data is significant and model-independent. 

Black hole binaries are known to have inter-band time lags above 10 keV (mostly from XTE data, \citealt{1999ApJ...510..874N,2014arXiv1402.4485G}), where a lag-energy plot shows a smooth increase with energy. Although their origin is not clear, they could be Comptonization lags \citep{1997ApJ...480..735K} or propagating fluctuations lags \citep{2001MNRAS.327..799K,2006MNRAS.367..801A}, with the latter being most likely. The key property in the Galactic black holes lags is their \emph{smoothness} with energy. The absence of any features in their lag-energy spectra linked them to the primary Comptonizing continuum. The power-law shape of this Comptonization component explains the smoothness of the lag-energy spectra in Galactic black holes \citep{2001MNRAS.327..799K}. The question is: are the lags presented here for \mcg similar to those in Galactic black holes?

Two points are considered in addressing the question. First, are the time-scales comparable assuming black hole systems in AGN are scaled-up version of Galactic black holes? Second: Do the lags have similar energy dependence?

The lags in black hole binaries are measured at frequencies of $\sim 1$ Hz for Cygnus-X1 as an example ($0.1-10$ Hz, \cite{1999ApJ...510..874N,2001MNRAS.327..799K}). The ratio of the masses between Cygnus-X1 and \mcg is $\sim10/2\times10^7 = 5\times10^{-7}$. Assuming the variability time-scale is proportional to the mass (both viscous and thermal time-scales scale with mass), a variability frequency of 1 Hz in Cygnus-X1 corresponds to a frequency of $5\times10^{-7}$ Hz in \mcg, which is smaller (i.e. time scale is longer) than probed with the current observations. Unlike the Galactic source time-scales, the \mcg observations probe time-scales that corresponds to thermal time-scales a few and at most tens of gravitational radii from the black hole, hinting at a different origin.

The second important point is the energy dependence. Lag-energy spectra in Galactic black hole are very smooth. \cite{2001MNRAS.327..799K} searched for any features in the lag-energy spectrum of Cygnus-X1 and ruled them out. The case of \mcg is different. It is not just the {\em NuSTAR} data that shows a feature around the 6 keV at the same frequencies, but more importantly, the much better quality combined data from XMM-{\em Newton} and {\em Suzaku} presented in section \ref{sec:ironK_rev} do also. It is by looking at the same-frequency lag-energy spectra below 10 keV that the lags above 10 keV can be properly interpreted.

Data below 10 keV strongly suggest a lag origin in relativistic reverberation. This comes from the existence of lags in the first place and then their energy dependence that resembles a relativistically-broadened iron line. In this picture, energies above 10 keV are also expected to produce a lag feature corresponding to the Compton hump from the same reflection spectrum as the iron line. As in the case of the iron line, the increase in the lag-energy spectrum above 10 keV is produced by an increase in the reflection contribution in the spectrum. Furthermore, the analysis presented in section \ref{sec:hump_lags} seems to indicate that there is a down-turn above 30 keV as expected after the Compton hump peaks and the reflection fraction starts to decrease again. Statistically, the turn-down is present at the $2\sigma$ level when considering the whole frequency band. The significance is higher for higher frequencies suggesting that the turn-down is real. The contribution of the intrinsic continuum lags remains however unclear.

Therefore, although continuum lags cannot be ruled out directly, the non-matching frequency scaling between Cygnus-X1 and \mcg, the non-smooth lag-energy spectra in \mcg compared to Cygnus-X1 and the possible down-turn at $\sim 30$ keV, together indicate that the lags above 10 keV in \mcg are different from those in Galactic black holes. Lags in Galactic black holes are continuum lags characterized by a smooth increase with energy. They are due to Comptonization or more likely due to propagating fluctuations. The lags in \mcg are different, and there are strong indications that they are due to relativistic reverberation.

The lag-energy spectrum of \mcg using the {\em NuSTAR} data at two frequency bins (Fig. \ref{fig:nu_lag_en}) also indicates a generally steeper spectrum at lower frequencies. As we have indicated, continuum lags inferred from a simple mass-scaling of those in Cygnus-X1 are expected at $10^{-7}-10^{-6}$ Hz, so there is a possibility they are contributing to the lowest frequencies in the bin giving a steeper profile. This means that lag-energy spectrum in this case is measuring the reflection fraction superimposed on top of a continuum lag that smoothly increase with energy. When the lowest frequencies containing the continuum are excluded from the frequency bin (orange squares in Fig. \ref{fig:nu_lag_en}-right), the general shape is now flatter, because it is only measuring the reverberation lags and the shape is essentially the reflection fraction only. This difference between low and high frequency energy-dependence is similar to what is seen below 10 keV in many objects \citep{2010MNRAS.401.2419Z,2011MNRAS.418.2642Z}. The case of Akn 564 for example \citep{2013MNRAS.434.1129K} shows clearly that the low frequency lags increase smoothly with energy, while at high frequencies relativistic reverberation traces the shape of a reflection component. With the Compton hump results in \mcg, this property of two lag processes dominating at different frequencies is further supported by including hard X-rays.

\section{Summary}
This work present three main results:
\begin{itemize}
\item Confirmation of iron K reverberation seen earlier \mcg using XMM-{\em Newton} data. New observations with XMM-{\em Newton}, {\em Suzaku} and {\em NuSTAR} confirm the existence of the lags.
\item Measurement the shape of the relativistic iron line at three time-scales through the calculation of lag-energy spectra. The shape of the line changes with time-scale. At long time scales, both the blue and red wings of the line are seen. At short time-scales, only the red wing is seen.
\item Using data from {\em NuSTAR}, we report the discovery of a time delay between energies $>10$ keV and the continuum. These lags are most likely due to reverberation in the reflection Compton hump.
\end{itemize}

\section*{Acknowledgment}
This work has been partly supported by NASA grant NNX14AF89G.
The research in this article has made use of data obtained from the {\em Suzaku} satellite, a collaborative mission between the space agencies of Japan (JAXA) and the USA (NASA). We made use of data from the {\em NuSTAR} mission, a project led by the California Institute of Technology, managed by the Jet Propulsion Laboratory, and funded by the National Aeronautics and Space Administration. Part of this work is based on observations obtained with XMM-{\em Newton}, an ESA science mission with instruments and contributions directly funded by ESA Member States and NASA.

\bibliographystyle{astron}
\bibliography{bibliography}

\end{document}